\documentclass[twocolumn,showpacs,aps,prb,superscriptaddress,floatfix]{revtex4}
\usepackage{graphicx}
\usepackage{dcolumn}
\usepackage{bm}

\begin{document}

\preprint{}

\newcommand{\nno}{NaNiO$_2$ }
\newcommand{\lno}{LiNiO$_2$ } 

\affiliation{Department of Physics and Astronomy, McMaster University,
Hamilton, Ontario, L8S 4M1, Canada}
\affiliation{Canadian Institute for Advanced Research, 180 Dundas St. W., 
Toronto, Ontario, M5G 1Z8, Canada} 

\author{M.J. Lewis}
\affiliation{Department of Physics and Astronomy, McMaster University,
Hamilton, Ontario, L8S 4M1, Canada}
\author{B.D. Gaulin}
\affiliation{Department of Physics and Astronomy, McMaster University,
Hamilton, Ontario, L8S 4M1, Canada}
\affiliation{Canadian Institute for Advanced Research, 180 Dundas St. W., 
Toronto, Ontario, M5G 1Z8, Canada} 
\author{L. Filion}
\affiliation{Department of Physics and Astronomy, McMaster University,
Hamilton, Ontario, L8S 4M1, Canada}
\author{C. Kallin}
\affiliation{Department of Physics and Astronomy, McMaster University,
Hamilton, Ontario, L8S 4M1, Canada}
\affiliation{Canadian Institute for Advanced Research, 180 Dundas St. W.,
Toronto, Ontario, M5G 1Z8, Canada}
\author{A.J. Berlinsky}
\affiliation{Department of Physics and Astronomy, McMaster University,
Hamilton, Ontario, L8S 4M1, Canada}
\affiliation{Canadian Institute for Advanced Research, 180 Dundas St. W.,
Toronto, Ontario, M5G 1Z8, Canada}
\author{H.A. Dabkowska}
\affiliation{Department of Physics and Astronomy, McMaster University,
Hamilton, Ontario, L8S 4M1, Canada}
\author{Y. Qiu}
\affiliation{National Institute of Standards and Technology, 100 Bureau 
Dr., Gaithersburg, MD, 20899-8562, U.S.A.} 
\affiliation{Department of Materials Science and Engineering, University 
of Maryland, College Park, MD, 20742, USA}
\author{J.R.D. Copley}
\affiliation{National Institute of Standards and Technology, 100 Bureau
Dr., Gaithersburg, MD, 20899-8562, U.S.A.}%

\title{Ordering and Spin Waves in \nno : A Stacked Quantum Ferromagnet} 

\begin{abstract} 
Neutron scattering measurements on
powder \nno reveal magnetic Bragg peaks and spin waves characteristic of
strongly correlated s=1/2 magnetic moments arranged in ferromagnetic
layers which are stacked antiferromagnetically.  This structure lends
itself to stacking sequence frustration in the presence of mixing between
nickel and alkali metal sites, possibly providing a natural explanation
for the enigmatic spin glass state of the isostructural compound, \lno.
\end{abstract} 

\maketitle 

\section{Introduction}

Low dimensional quantum magnets are well appreciated as fertile ground for spin liquid and other exotic quantum
states of matter\cite{reviews}. \nno and \lno are isostructural nickel-based quantum magnets with layered
triangular structures.  The enigmatic magnetic phase behaviour associated with \lno has been the subject of
speculation for two decades \cite{Hirakawa,Reimers, Yoshizawa, Hirota, Kemp, Kitaoka, Reynaud, Mostovoy}.  This
speculation has been largely fueled by the absence of observed transitions to long range magnetic and orbital
order in \lno.  In contrast \nno is known to undergo a Jahn-Teller 
structural distortion at high temperatures
signalling orbital order\cite{Chappel}, and a low temperature anomaly in its susceptibility is often associated 
with
long range antiferromagnetic order\cite{Chappelmag}.  However, like \lno, no definitive magnetic neutron
scattering signature of such an ordered state has been observed in \nno until now.  We report the
observation of magnetic Bragg peaks and spin wave scattering in \nno, which establishes its relatively simple
magnetic structure and which may shed light on why \lno has difficulty finding an ordered state at low temperatures.

Both \nno and \lno are comprised of stackings of triangular layers which
alternate between NiO and AO, where A is either Li$^+$ or Na$^+$, as shown
schematically in Fig. 1.  The distance between neighboring triangular
layers of magnetic Ni along the c-direction is about 5.2 \AA.  Most of
the extensive preceding work associated with the magnetism in these 
materials,
assumes a low spin Ni$^{3+}$, s=1/2 magnetic moment.  However some
discussion of nickel in the s=1 Ni$^{2+}$ oxidation state bound to an
s=1/2 hole on a neighbouring O$^{-}$ ion, resulting in an s=1/2 Zhang-Rice
singlet has been offered \cite{Sawatsky, Mizokawa}.  In both scenarios,
either the nickel spin itself, or the composite NiO spin is an extreme
s=1/2 quantum mechanical entity, and a saturation magnetization
corresponding to 1 $\mu_B$ per formula unit is observed in both \nno and
\lno\cite{Holzapfel}.

\begin{figure}
\centering
\includegraphics[width=6.5cm]{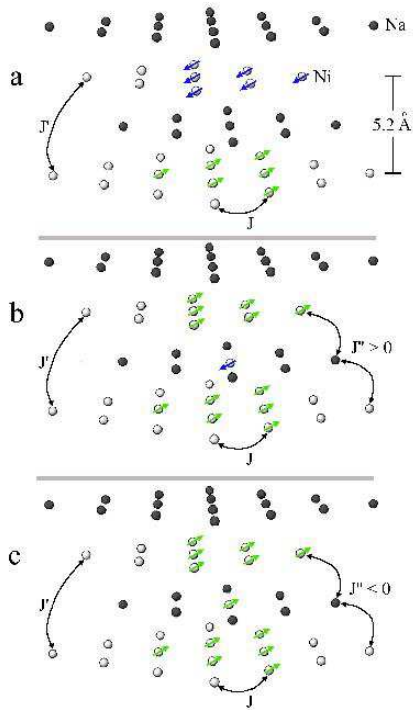}
\caption{(a) The schematic magnetic structure of \nno below T$_N$ $\sim$ 
23 
K.  
(b) Stacking sequence frustration 
induced by impurity Ni spins on the alkali metal sublattice
for antiferromagnetic (J$^{\prime\prime}$ $>$ 0) coupling.  (c) The 
same stacking sequence is favored by  
ferromagnetic 
(J$^{\prime\prime}$ $<$ 0, bottom) coupling.} 
\label{Figure 1} 
\end{figure}

The possibility of antiferromagnetic coupling between these quantum
moments within the triangular plane has generated much interest in the
materials.  The potential for geometrical frustration and exotic quantum
ground states under such conditions have been well appreciated since
Anderson \cite{Anderson} suggested a collective singlet, resonating
valence bond ground state without long range order for such a system.  
Indeed this possibility figures prominantly in the
discussion of both early and more recent experimental work on 
\lno\cite{Hirakawa, Kitaoka}.  More
generally, it is well known that the combination of antiferromagnetism and
certain lattice symmetries based on triangles and tetrahedra leads to
phenomena known broadly as geometrical frustration\cite{frus}.

As mentioned above, the possibility of such exotic behaviour has been
motivated, in part, by the absence of direct experimental signatures of
conventional magnetic ordering, such as the observation of magnetic Bragg
peaks by neutron diffraction. Given the difficulties associated with the
study of these materials, it is small wonder that progress has been slow.  
To date they have been available only as polycrystalline materials, and
with a single s=1/2 magnetic moment per formula unit, the moment density
is low.  In addition, \lno contains Li, whose strong neutron absorbtion is
problematic for neutron scattering measurements.  Finally, it is known 
that
the similarity in ionic radii between Li$^+$ and Ni$^{3+}$, both 
$\sim$ 0.7 \AA, leads to mixing between these two
sublattices\cite{Reimers}, further complicating the magnetic behaviour of
\lno. With an ionic radius of $\sim$ 1 \AA, Na$^+$ is a much larger ion,
which makes such mixing unlikely.  For these latter two reasons, one
expects \nno to be more amenable to a neutron scattering study than \lno.

\section{Materials and Experimental Methods}

\nno crystallizes into the rhombohedral space group R3m at high
temperatures, before undergoing a cooperative Jahn-Teller distortion
leading to a structure with the space group C2/m below $\sim$ 480 K
\cite{Chappel,Dick}. This structural phase transition lifts the orbital
degeneracy between the $\mid$3z$^2$-r$^2$$>$ and 
$\mid$x$^2$-y$^2$$>$ states 
within the
e$_g$ doublet in \nno \cite{Chappel}. The resulting room temperature structure\cite{Chappel} 
in \nno is characterized by lattice parameters a=5.31 \AA, b=2.84 \AA, c=5.57 \AA, and
$\beta$=110.44 degrees, and the structure no longer displays edge-shared, equilatoral
triangles within the a-b plane.

A similar structural distortion does
not obviously occur in \lno although anomolies in the susceptibility are
reported near 480 K in \lno \cite{Reynaud}, which may be related to a {\it
local} structural distortion that cannot attain long range order due
to mixing of the Li and Ni sublattices.  

The lack of obvious orbital ordering in \lno and the orbital degeneracy
which would arise in its absence has motivated many theoretical authors to
consider coupled orbital and spin degrees of freedom\cite{Kitaoka,
Reynaud,Mostovoy,Feiner,Li,Bossche,Vernay} as a mechanism to suppress
magnetic ordering at low temperatures.  Other recent work suggests orbital 
and spin degrees of freedom are decoupled\cite{Holzapfel}.

\begin{figure}
\centering
\includegraphics[width=8.5cm]{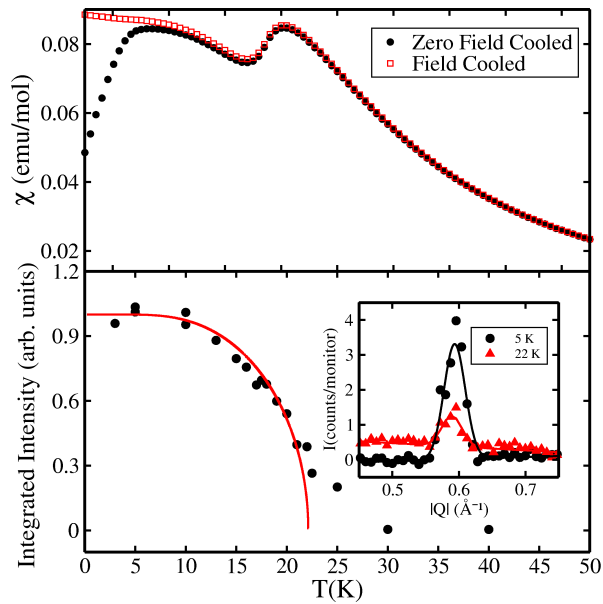}
\caption{top panel: Field-cooled and zero field cooled susceptibilities of 
\nno are shown.  Bottom panel: The temperature dependence of the 
integrated intensity of the (0,0,1/2) magnetic Bragg peak at 
Q$\sim$ 0.6 $\mbox{\AA} ^{-1}$ is shown, indicating a continuous transition near 
T$_N$ 
$\sim$ 23 K.  The inset shows net scattering at two different 
temperatures below T$_N$ as described in the text} 
\label{Figure 2.}
\end{figure}

Stoichiometric amounts of Na$_2$O$_2$ and NiO were mixed and pelletized
in an Ar atmosphere. These pellets were subsequently
annealed at 973 K in O$_2$ for 70 hours with one intermediate grinding. We
prepared a 30 gram polycrystalline sample of \nno in this manner and
characterized a small part of it with SQUID magnetic susceptibility
techniques. The resulting field-cooled (FC) and zero field cooled (ZFC)  
dc susceptibilities are shown in the top panel of Fig. 2.  The magnetic
phase transition near T$_N$$\sim$ 23 K is immediately clear as a peak in
$\chi$(T).  In addition, a break between the FC and ZFC susceptibilities
below $\sim$ 10 K is seen, indicating some glassiness within the ordered
state at these low temperatures.

Time-of-flight neutron scattering measurements were performed on this 30
gram sample of \nno using the Disk Chopper Spectrometer (DCS) at the NIST
Center for Neutron Research.  The DCS uses choppers to create pulses of
monochromatic neutrons whose energy transfers on scattering are determined
from their arrival times in the instrument's 913 detectors located at
scattering angles from -30 to 140 degrees. Measurements were performed
with 3.2 and 5.5 \AA incident neutrons.  Using 5.5 \AA incident 
neutrons,
the energy resolution was ~0.075 meV\cite{Cook}. 

\section{Experimental Results}

Typical data sets at low 
temperature
are shown in Fig. 3a, b and c, with the intensity scale chosen to
highlight detail in the inelastic scattering.  Cuts through the elastic
scattering (integrating in $\hbar \omega$ between $\pm$ 0.1 meV) of 
this data are shown in the inset to the lower panel of Fig. 2.  Here 
we clearly see the appearace of new Bragg peaks below T$_N$ $\sim$ 23 
K.   These particular data sets correspond to the subtraction of a high 
temperature, 30 K, data set from data sets at the indicated low 
temperatures and are centered around the Q=0.6 $\mbox{\AA} ^{-1}$ region of 
reciprocal space, the position at which the lowest-Q Bragg peak appears.

\begin{figure}
\centering
\includegraphics[width=8.4cm]{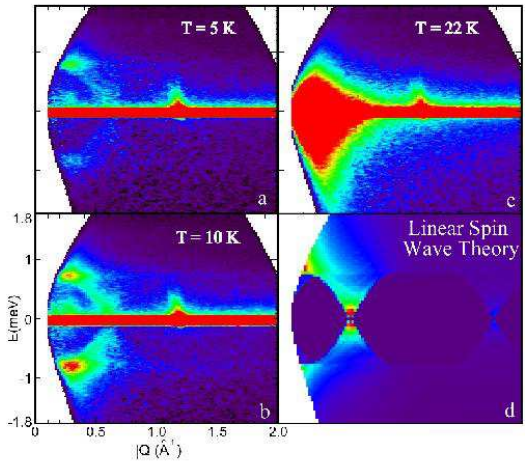}
\caption{Maps of neutron scattering from \nno at three temperatures near 
and below T$_N$ $\sim$ 23 K; at a) 5 K, b) 10 K, and c) 22 
K.  The bottom right panel, d, shows the results of a linear spin 
wave theory calculation, at 10 K which can be compared directly to the 
experiments. The linear intensity scale has been chosen to highlight the 
inelastic spin wave scattering.} 
\label{Figure 3} 
\end{figure}

This and several other very weak magnetic Bragg peaks ($\sim$ 10$^{-3}$ of 
the strongest
nuclear Bragg peaks) are evident and can be accounted for both in position
and relative intensity by a simple magnetic structure in which the s=1/2
moments on the nickel sites are arranged in ferromagnetic sheets within
the triangular planes, and the sheets are antiferromagnetically stacked
along c, such that moments on neighbouring planes are $\pi$ out of phase
with each other. The details of the modeling of this elastic magnetic 
scattering are consistent with a rather large ordered moment, close to 
1 $\mu_B$/Ni, and do not support composite Zhang-Rice singlet 
magnetic moments associated with NiO.  

The Bragg peak shown in the bottom panel of Fig. 2 is indexed as (0,0,1/2)  
and arises due to a $\pi$ phase shift across the 5.2 \AA between
triangular Ni planes, thereby appearing at Q=$\pi$/5.2 \AA $\sim$ 0.6 
$\mbox{\AA}^{-1}$.  The integrated intensity of this Bragg peak is shown in the
bottom panel of Fig. 2.  Both this intensity and that of other magnetic
Bragg peaks tend to zero near T$_N$$\sim$ 23 K,
consistent with a continuous phase transition.  The rounded nature of the 
order parameter near T$_N$ is unusual, and possibily due to dimensional 
cross over.

The details of the magnetic structure, to be presented elsewhere, are
determined by the relative intensities of the measured magnetic Bragg
peaks at low temperatures.  Here we simply note that the form of the
magnetic neutron scattering cross section requires that Bragg intensity at
(0,0,1/2) be due to components of ordered moment within the triangular
plane, and that the relative intensities require that the moments make a 
relatively large angle with respect to the triangular plane.

\begin{figure}
\centering
\includegraphics[width=8.5cm]{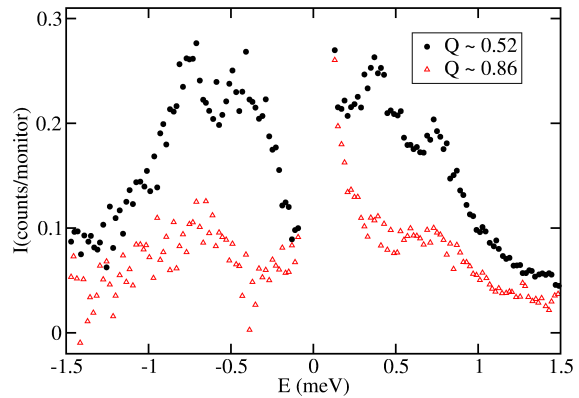}
\caption{Cuts through the maps of scattering at 10 K shown in Fig. 3b.
These data approximate constant-Q scans at Q=0.52
$\mbox{\AA}^{-1}$ and 0.86 $\mbox{\AA}^{-1}$ and clearly show a largely dispersionless
excitation
at
$\hbar\omega$ $\sim$ 0.7 meV, in addition to the Goldstone mode.}
\label{Figure 4}
\end{figure}

The inelastic scattering in Fig. 3 clearly shows spin wave excitations
going to zero energy at the (0,0,1/2) magnetic zone centre, and reaching a
zone boundary energy of $\sim$ 0.7 meV at Q$\sim$0.3 $\mbox{\AA}^{-1}$.  In
addition, rather weak but clearly observable inelastic scattering is seen
in a dispersionless inelastic feature at $\sim$ 0.7 meV which extends
across all Q measured.  This is most clearly seen in cuts through this
data, which approximate constant-Q scans, and which are shown for two
different Q's in Fig. 4.  All of these excitations can be seen on both the
neutron energy loss (+ve side) and neutron energy gain (-ve side) of zero
energy transfer, as required by detailed balance.  The inelastic data in
Fig. 3, has been corrected for detector efficiency only.

\begin{figure}
\centering
\includegraphics[width=8.4cm]{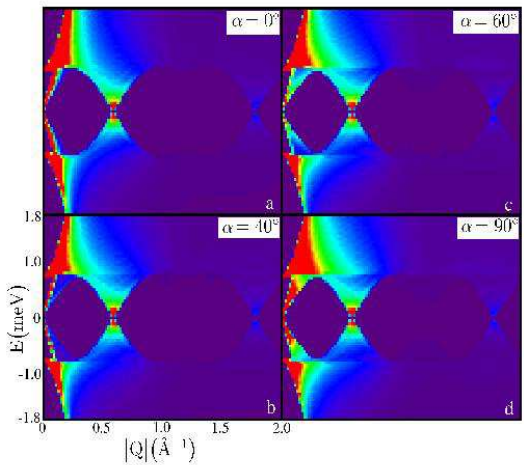}
\caption{Maps of calculated neutron scattering from \nno using linear spin
wave theory is shown at 10 K, and for four different easy plane orientations
relative to the triangular basal plane.  $\alpha$=0, shown in a),  
corresponds 
to a coincidence between the easy plane and the triangular plane, while 
$\alpha$=90, in d), corresponds
to these two planes being normal to each other.  The calculation for 
$\alpha$=40 degrees in
panel b) is the same as that shown in Fig. 3d).  These calculations show 
that a relatively large angle $\alpha$ is required between the magnetic easy plane 
and the triangular basal plane, in 
order for the dispersionless excitation near 0.7 meV to be observable.}
\label{Figure 5}
\end{figure}

The inelastic scattering meaurements near T$_N$ show the spin wave band 
along c$^*$ to soften completely, such that the envelope of the spin wave 
dispersion fills in with inelastic intensity, as shown in Fig. 3c.  These 
measurements indicate the nature of the transition at T$_N$ $\sim$ 23 K is 
a loss of registry between well-correlated ferromagnetic triangular 
planes.  This likely also explains the previous observation of small angle 
neutron scattering in \lno\cite{Yoshizawa} as originating from the 
collapse of the inelastic spin wave scattering, which occurs below Q$\sim$0.6 
$\mbox{\AA}^{-1}$. Such inelastic scattering would be integrated up 
in a diffraction, SANS experiment producing a temperature-dependent 
signal at small-Q, whose intensity peaks near T$_N$.

\section{Linear Spin Wave Theory and Comparison to Experiment}

We have also carried out linear spin wave theory calculations appropriate
to s=1/2 moments interacting on the Ni$^{3+}$ sublattice of \nno , with
the following microscopic Hamiltonian:

\begin{eqnarray}
\mathcal{H} & = & J\sum_{ab_{nn}}{\bf S_{i}S_{j}} + K\sum_{ab_{nn}}{S_{i}^zS_{j}^z} + 
J'\sum_{c_{nn}}{\bf{S_{i}S_{j}}}
\end{eqnarray}

where the exchange integral within the triangular {\it ab} basal plane is 
ferromagnetic, and relatively strong, J=-2.5 meV, and that along the 
stacking {\it c} direction is antiferromagnetic and relatively weak, J$^\prime$=+0.16 meV.
In addition we have included an anisotopy term, K, whose strength is set 
equal to J$^\prime$, and whose purpose is to preferentially restrict the spins 
to a plane perpendicular to z.  All interactions are nearest neighbour
only.  The inclusion of this anisotropy term is phenomenological, producing two 
transverse branches in the spin wave spectrum along c$^*$, as is observed experimentally. 
Note that the Dzyaloshinski-Moriya interaction, which requires a lack of inversion symmetry
between the sites whose coupling it mediates, is not allow by the space group of \nno.
Hence anisotropic exchange is expected to be the leading order anisotropic interaction.

Im$\chi$({\bf Q}, $\hbar \omega$) for such a system was calculated within
linear spin wave theory at T=10 K and this result was angularly averaged
to produce theoretical expectations appropriate to an inelastic neutron
scattering study carried out on a polycrystalline sample.  This is shown
in Fig. 3d and in all the panel of Fig. 5. 

This calculation employs a z-direction in Eq. 1, normal to the easy plane, 
which defines the easy plane.  In general, the resulting magnetic easy plane 
makes some angle, $\alpha$, with respect to the triangular basal plane.  
Therefore the magnetic easy plane is not neccesarily coincident with the triangular plane.  
The calculation shown in Fig. 3d employs an angle of $\alpha$=40 degrees.  
Fig. 5 shows the results of this calculation for the expected 
powder-averaged inelastic neutron scattering using four different values of $\alpha$:
a) $\alpha$=0 degrees (easy plane coincident with 
triangular plane), b)
$\alpha$=40 degrees, c) $\alpha$=60 degrees, and d) $\alpha$=90 degrees (easy plane normal to
the triangular plane).

As can be seen, a non-zero $\alpha$ is required
in order for the dispersionless mode at $\sim$ 0.7 meV to have observable
weight. If the magnetic easy plane was coincident with the triangular
basal plane, as they are at $\alpha$=0, fluctuations out of this plane would be along
c$^*$, and not observable along this direction, by virtue of the polarization dependence of the neutron scattering 
cross section. In this $\alpha$=0 case, only the Goldstone mode is predicted to be observed.  

As can be seen comparing Fig. 3b and 3d, the quantitative
agreement between the calculation of the spin wave spectrum and the
measurements at 10 K is very good, although some minor discrepancies are
evident.  The measured spin wave excitations are identified as
corresponding to those propagating along the stacking direction.
Furthermore, the zone boundary spin wave energy, at Q=0.3 $\mbox{\AA}^{-1}$, is given by $\Delta _{ZB} =
6S\sqrt{{({J^\prime}^2+J^\prime K)}}$ which reduces to $\Delta _{ZB} = 
6\sqrt{2}SJ^{\prime}$ for the case where
J$^\prime$=K.  Reading $\Delta _{ZB}$ =0.7 meV off the data, we get J$^\prime$=0.16 meV. This direct measurement
of J$^\prime$ can be combined with the measured $\Theta _{CW}$$\sim$3.5 meV obtained from fits to the high
temperature susceptibility \cite{Chappelmag,Kemp} to produce an estimate for the stronger in-plane ferromagnetic
exchange.  Using $k_B\Theta _{CW}=-{{6S(S+1)}\over 9}(3(J+J^\prime)+K)$ we obtain J=-2.5 meV.
 
\section{Discussion and Conclusions}

Our measurements of the spin structure and dynamics in \nno directly
reveal a rather simple magnetic structure corresponding to an
antiferromagnetic stacking of ferromagnetic sheets.  Such a structure 
is very sensitive to frustration of the stacking
sequence, and therefore frustration in the ability of the structure to
attain true three dimesional long range order, due to mixing of the alkali
metal and transition metal sublattices.  This is precisely the nature of
chemical disorder shown to be relevant in \lno at the 1-3 $\%$ level 
\cite{Reimers, Nunez}.

This scenario is illustrated schematically in Fig. 1.  This 
drawing shows 3 sets of interacting ferromagnetic layers of Ni 
moments.  Figure 1(a) shows the simple antiferromagnetic 
stacking of ferromagnetic layers determined from our neutron 
measurements.   Figure 1 (b) and (c) show the consequences of impurity 
spins which couple either 
antiferromagnetically (b) or ferromagnetically (c) to  
neighboring Ni layers.  It is clear that misplaced magnetic
Ni ions sitting in the alkali metal triangular planes provide an exchange 
pathway along the stacking
direction which will frustrate the stacking sequence {\it independent of
whether these impurity spins couple
ferromagnetically or antiferromagnetically to the moments on Ni triangular
planes above and below.} This is due to the fact that {\it two} such
pathways link neighbouring Ni triangular planes, thus the sign of this
interaction is irrelevant.  Furthermore, as these impurity nickel spins
are a factor of two closer along the stacking direction to spins within
the Ni triangular planes than those residing within the Ni triangular
planes, the magnitude of this frustrating Ni - impurity Ni - Ni
interaction is expected to be relatively strong.  

If, as has been argued\cite{Reynaud, Holzapfel}, orbital degeneracy is not
responsible for it, a natural and simple explanation for the spin glass
phase in \lno ensues as a consequence of this structure.  Samples of \lno
display weak mixing between the Ni and Li sublattices, providing a sample
dependent tuning of the strength of the stacking sequence disorder.  This
leads to a glass transition which depends on the precise level of disorder
in the sample and which will occur at temperatures on the order of that
associated with phase coherence between strongly correlated ferromagnetic
layers, ~ T$_N \sim$ 23 K.  This scenario is qualitatively similar to the
ferrimagnetic clusters previously proposed\cite{Nunez, Chappel2} to
explain the spin glass phase in \lno, with the exception that the stacking
sequence frustration occurs independent of the sign of the coupling
between impurity spins and the magnetic layers.

To conclude, new elastic and inelastic neutron scattering measurements
have directly determined the simple antiferromagnetic structure of \nno
below T$_N$ $\sim$ 23 K.  These measurements allow a microscopic understanding
of the magnetically-ordered state in \nno and may
provide the basis for a simple explanation of the phase behavior exhibited
in \lno.

\section{Acknowledgements}

We wish to acknowledge useful contributions from I. Crooks, J. van Duijn,
S-H. Lee, S. Park, and G. Sawatzky. This work was supported by NSERC of
Canada, and utilized facilities supported in part by the NSF under
Agreements DMR-9986442 and DMR-0086210.


%
%






\end{document}